# What is Learning Analytics about? A Survey of Different Methods Used in 2013-2015

**Mohammad Khalil**
*Graz University of Technology, Graz, Austria*

**Martin Ebner**
*Graz University of Technology, Graz, Austria*

**Abstract**

*The area of Learning Analytics has developed enormously since the first International Conference on Learning Analytics and Knowledge (LAK) in 2011. It is a field that combines different disciplines such as computer science, statistics, psychology and pedagogy to achieve its intended objectives. The main goals illustrate in creating convenient interventions on learning as well as its environment and the final optimization about learning domain's stakeholders (Khalil & Ebner, 2015b). Because the field matures and is now adapted in diverse educational settings, we believe there is a pressing need to list its own research methods and specify its objectives and dilemmas.*

*This paper surveys publications from Learning Analytics and Knowledge conference from 2013 to 2015 and lists the significant research areas in this sphere. We consider the method profile and classify them into seven different categories with a brief description on each. Furthermore, we show the most cited method categories using Google scholar. Finally, the authors raise the challenges and constraints that affect its ethical approach through the meta-analysis study.*

*It is believed that this paper will help researchers to identify the common methods used in Learning Analytics, and it will assist by establishing a future forecast towards new research work taking into account the privacy and ethical issues of this strongly emerged field.*

## Introduction

Since the first Learning Analytics and Knowledge (LAK) conference in 2011 as well as the Horizon Report in 2013 (Johnson et al., 2013), learning analytics is considered to be an emerging field that would be applied in the different educational settings. This field provides tools and technologies that offer the potentials to do proper interventions and improve education in general. The Society for Learning Analytics and Research (SoLAR) defined it as "the measurement, collection, analysis and reporting of data about learners and their contexts, for purposes of understanding and optimizing learning and the environments in which it occurs". Several studies exchanged views about learning analytics goals. For instance, Khalil and Ebner (2015) introduced learning analytics lifecycle and listed the main surveyed objectives of the past four years of the LAK conferences.

They listed interventions, predictions, reflection, awareness, personalization, recommendation and benchmarking as the main goals. These goals conformed to Siemen's defined techniques back in 2012 (Siemens, 2012). In addition to that, different frameworks have been introduced to define both key objectives and dilemmas of this field. In his paper "The Learning Analytics Cycle: Closing the loop effectively", Clow argued that a successful learning analytics should be winding up into feeding back the product to learners in order to make effective intervention(s) (Clow, 2012). Whiles Ferguson indexed remarkable challenges of ethics, distinct perspectives from stakeholders' field of vision and the methods to use in order to make these goals achievable (Ferguson, 2012).

For the time being, there is a large variety of educational environments such as MOOC-platforms, LMS, virtual environments, etc. These educational information systems hold "Big Data" of learners that create huge data repositories. According to learning analytics definition, the data need to be analysed by typical methodologies in order to reflect benefits on learning and teaching. The beginning of lively discussions on the differences between learning analytics and educational data mining were mainly residing to the opposing opinions of using tools and methodologies in both fields (Baker et al., 2012). Nevertheless, educational data mining and learning analytics are enriched by the methods of data mining and analytics in general (Baker & Siemens, 2013). In its first stages, researchers of learning analytics frameworks and structure were discussing methods such as visualizations, data mining techniques (Elias, 2011), social network analysis (Ferguson, 2012), and sentiment analysis (Siemens, 2012), in addition to statistics which was also mentioned as a required tool to build learning prediction models (Campbell, DeBlois & Oblinger, 2007).

SoLAR brought to success the annual organization of LAK conferences since 2011. Accordingly, several categories of methods to analyse educational datasets were used. Most of these methods tend to process data quantitatively and qualitatively to discover interesting hidden patterns. Baker and Siemens (2013) mentioned that educational data is what drives new methods to be used in learning analytics. They said: *"The specific characteristics of educational data have resulted in different methods playing a prominent role in EDM/LA than in data mining in general, or have resulted in adaptations to existing psychometrics methods"*. In this paper, we survey publications from LAK conference from 2013 to 2015. The purpose is to list the most common methods used in the field of learning analytics in the last three years. We believe this paper provides different benefits because learning analytics becomes an important field by itself and is now completely matured into being adapted in different educational institutions and applications. The main advantages are:

1. It helps learning analytics researchers to identify common methods in use in order to reach intended goals.

2. It determines methods that are highly cited, e.g. by Google scholar (http://scholar.google.com), and establish a future forecast towards new research work.

3. Finally, it assists to compare the beginning view about learning analytics methods and the on-going current version.

In addition, the paper aims to guide future researchers into further advances in this field and meets the "Smart Learning Excellence" theme of Innovation Arabia 9 conference which is "The next wave of innovations in Smart-Learning" that mainly considers Big Data and Learning Analytics as a new wave in educational technology.

We have organized this paper into the following sections. First, we list the methodology employed to extract methods used in learning analytics publications. We then show statistical data and describe methods in detail with remarks about their types. Finally, we discuss and summarize the main conclusions and list the constraints and the ethical issues of learning analytics.

**Research Design**

As mentioned before, the conference of learning analytics and knowledge is considered to be the first and the largest repository of learning analytics publications. We mainly focused on it, and surveyed 91 papers from LAK 2013 (Suthers et al., 2013), LAK 2014 (Pistilli, Willis & Koch, 2014), and LAK 2015 (Baron, Lynch & Maziarz, 2015), with an additional supplementary literature from other sources. We excluded papers with topics philosophy, frameworks and conceptual studies of learning analytics for the reason that they address structures and do not accommodate a mechanism for revealing patterns. We also faced papers with unclear methods, and these were excluded too. At the end, 78 publications were set for examination. This study was influenced by the work of Romero and Ventura (Romero & Ventura, 2010), and Dawson et al. (Dawson et al., 2014). The classification of methods was based on reading the abstract, keywords, general terms, methodology section and the conclusion of each paper. In some publications, we paid more analysis into examining literature and the reference list. Furthermore, we collected the total number of citations for each analysed paper from Google scholar and observed the trending topics.

**Learning Analytics Methods**

Learning analytics is a combination of different disciplines like computer science, statistics, psychology, and education. As a result, we realized different analysis methods that do not only tend to be too technical but rather pedagogical. Before classifying the analysis methods, we have been gravitated towards the beginning topics of the emergence of learning analytics, which briefly described methods and tools for collecting data and analyzing them (Ferguson, 2012; Siemens, 2012). However, the survey reveals more methods being used to examine learners' data. Our main methods categories, which will be explained in detail in section 3.1, are: (a) data mining techniques; (b) statistics and mathematics; (c) text mining, semantics and linguistic analysis; (d) visualization; (e) social network analysis; (f) qualitative analysis; and (g) gamification. Figure 1 shows grouping of the methods used in learning analytics for LAK publications with the number of papers in each category. It should be noted that some publications might be referenced in a different category. Moreover, a paper could be referenced in multiple methods category.

The bar plot in the figure shows that researchers of 31 publications used data mining techniques and 26 research studies used statistics and mathematics to analyse their data. This makes both of these two methods as the top most employed techniques of analysis. We also see that "Text Mining, Semantics and Linguistic" analyses as well as visualizations are being used in 13 LAK publications equally. However, social network analysis and qualitative analysis as well as gamification were the least used techniques.

**Figure 1. Number of the examined LAK papers grouped by methods. Some papers share more than single category**

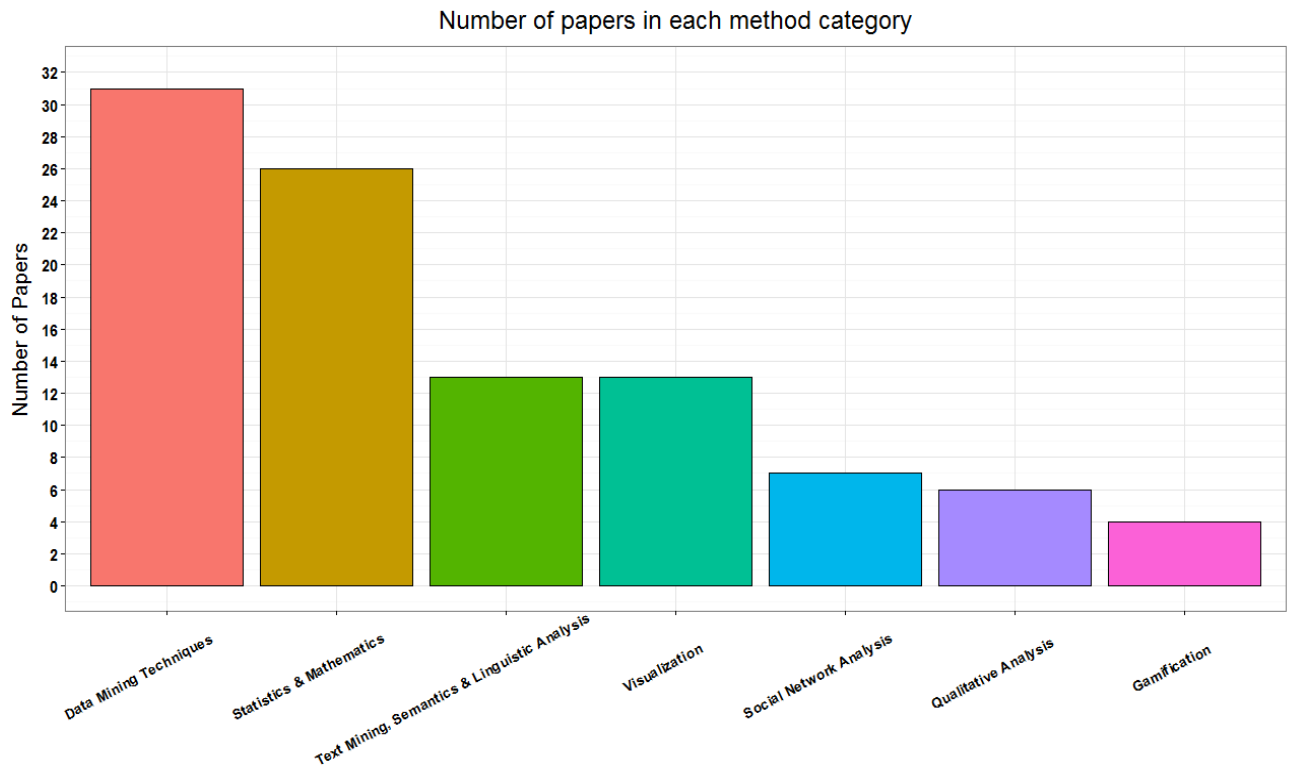

*Categories*

This section lists the methods categories in more detail and states relevant publications under each method.

***Data Mining Techniques***: Data mining tends to make sense out of data. The definition of learning analytics cited the similar idea, namely understanding the data, but in this case, about learners. The survey shows that data mining techniques are the most used method for analyzing and interpreting the learners' log data. The decision tree algorithm was used to predict the performance drop and the final outcome of students in a Virtual Learning Environment (VLE) (Wolff et al., 2013). Other researchers used several classification techniques such as step regression, Naive Bayes and REP-Trees to study students' behavior and detect learners who game the system (Pardos et al., 2013). While clustering was used to propose an approach for the purposes of enhancing educational process mining based on the collected data from logs and detecting students at risk (Bogarín et al., 2014). Discovering relations between two factors was observed by using multiple linear regression analysis to forecast the relation between studying time and learning performance (Jo, Kim & Yoon, 2014). Moreover, data mining is used for assessment such as the work at the University of Missouri-Columbia, which proposed an automated tool to enable teachers assess students in online environments (Xing, Wadholm & Goggins, 2014). It was remarked that regression analysis was the common mechanism among data mining techniques.

***Statistics and Mathematics***: Statistics is the science of measuring, controlling, communicating and understanding the data (Davidian & Louis, 2012). Publications show that researchers have been using descriptive statistics and mathematics, such as the mean, median and standard deviation to signify their results. In addition, inferential statistics was used side by side with data mining in some of the publications. Markov chain was used to study school students behavior in solving multiplication (Taraghi et al., 2014). Different statistical techniques were operated to build a grading system (Vogelsang & Ruppertz, 2015). Additionally, statistical discourse analysis with Markov chain was employed to study online discussions and summarize demographics (Chiu & Fujita, 2014), as well as examining student problem solving behavior and adapting it into tutoring systems (Eagle et al., 2015).

***Text mining, Semantics & Linguistic Analysis*:** Publications which refer to ontologies, mining texts, discourse analysis, Natural Language Processing, or study of languages are set to be in this category. Some studies refer to text analysis for assessment purposes of short answer questions (Leeman-Munk, Wiebe & Lester, 2014), or to enhance collaborative writing between students (Southavilay et al., 2013). Contextualizing user interactions based on ontologies to illustrate a learning analytics approach (Renzel & Klamma, 2013). Linguistic analysis was clearly used in parsing posts from students for prediction purposes (Joksimović et al., 2015). Finally, online discussion forums were analysed to pioneer an automatic dialogue detection system in order to develop a self-training approach (Ferguson et al., 2013).

***Visualization:*** When the information is visually presented to the field experts, efficient human capabilities rise to perceive and process the data (Kapler & Wright, 2015). Visual representations take the advantages into expanding human decisions within a large amount of information at once (Romero & Ventura, 2010). There are several studies that cited visualization as a method to analyse the data and deliver information to end users, such as: building a student explorer screen to prepare meetings and identifying at-risk students by the teachers (Aguilar, Lonn & Teasley, 2014). Studying MOOC's attrition rate and learners' activities (Santos et al., 2014); Building an awareness tool for teachers and learners (Martinez-Maldonado et al., 2015), and a dashboard for self-reflection goals (Santos et al., 2013). Information can be interpreted into heat maps, scatterplots, diagrams, and flowcharts which were observed in most of the statistical, mathematical and data mining based publications.

***Social Network Analysis*:** Abbreviated as SNA. It focuses on relationships between entities. In learning analytics, SNA can be used to promote collaborative learning and investigate connections between learners, teachers and resources (Ferguson, 2012). Moreover, it can be employed in learning environments to examine relationships of strong or weak ties (Khalil & Ebner, 2015). This category includes network analysis in general and Social Learning Analytics (SLA). The survey observed researchers who: built a collaborative learning environment by visualizing relationships between students about the same topic (Schreurs et al., 2013). A two-mode network was used to study students' patterns and to classify them into particular groups (Hecking, Ziebarth & Hoppe, 2014). It was also used with a grading system in a PLE to examine the centrality of students and grades (Koulocheri & Xenos, 2013). Again, not so far from this survey study, a network analysis was done to analyse citations of LAK conference papers (Dawson et al., 2014). The authors studied the degree centrality and pointed out the emergence and isolated disciplines in learning analytics. SNA was used to analyse data of connectivist MOOCs by examining interactions of learners from social media websites (Joksimović et al., 2015).

**Figure 2. Number of Google scholar citations of the examined LAK papers based on methods category. Retrieved on 26$^{th}$ October, 2015.**

*Qualitative Analysis:* This category is related to the decisions based on explained descriptions of the analysts. For instance: i) a qualitative evaluation with data mining techniques was made to understand the nature of discussion forums of MOOCs (Ezen-Can et al., 2015); ii) Usage of qualitative interviews, which are answered by words to build a learning analytics module of understanding fractions for school children (Mendiburo, Sulcer & Hasselbring, 2014); iii) Qualitative meta-analysis to investigate teachers' needs in technology enhanced learning covered by the umbrella of learning analytics (Dyckhoff et al., 2013).

*Gamification:* It is the use of game mechanics and tools to make learning and instruction attractive and fun (Kapp, 2012). This method is considered as a technique on its own because of its relevant appearance in educational workshops and the requests to make learning entertaining. Some examples are using rewards points and progress bar to enhance the retention rate and building a gamified grading system (Holman, Aguilar & Fishman, 2013), or presenting a competency map with progress bars, pie charts, labels and hints to improve students' performance (Grann & Bushway, 2014). A significant study on monitoring students in a 3D immersive environment was also advised as another type of gamification techniques (Camilleri et al., 2013).

**Prominent Methods and Discussion**

In this section, we consider learning analytics methods that have been frequently cited. We used the Google scholar as a foundation to check methods' popularity. All the data were collected recently and retrieved before the submission date. Figure 2 shows Google scholar citations for the analysed LAK conference papers based on the methods category. The publications with the method type Data Mining and Techniques, were the most cited articles (452 citations). The ultimate number of citations in this survey belongs to the paper of (Kizilcec, Piech & Schneider, 2013) with 236 citations. Although we took into consideration the time span of publications, we see that articles that belong to MOOCs are the most cited papers. Statistics and Mathematics publications were cited 363 times. Qualitative analysis and gamification publications were the least cited articles. In figure 3, we show a density plot of publications' citations grouped by year. The x-axis records number of citations converted into logarithmic scale to ease the reading. The y-axis records the density of publications per year. Since we did not survey a fair number of publications per year, we intended to use this plot instead of histogram plot, which is highly sensitive to bin size.

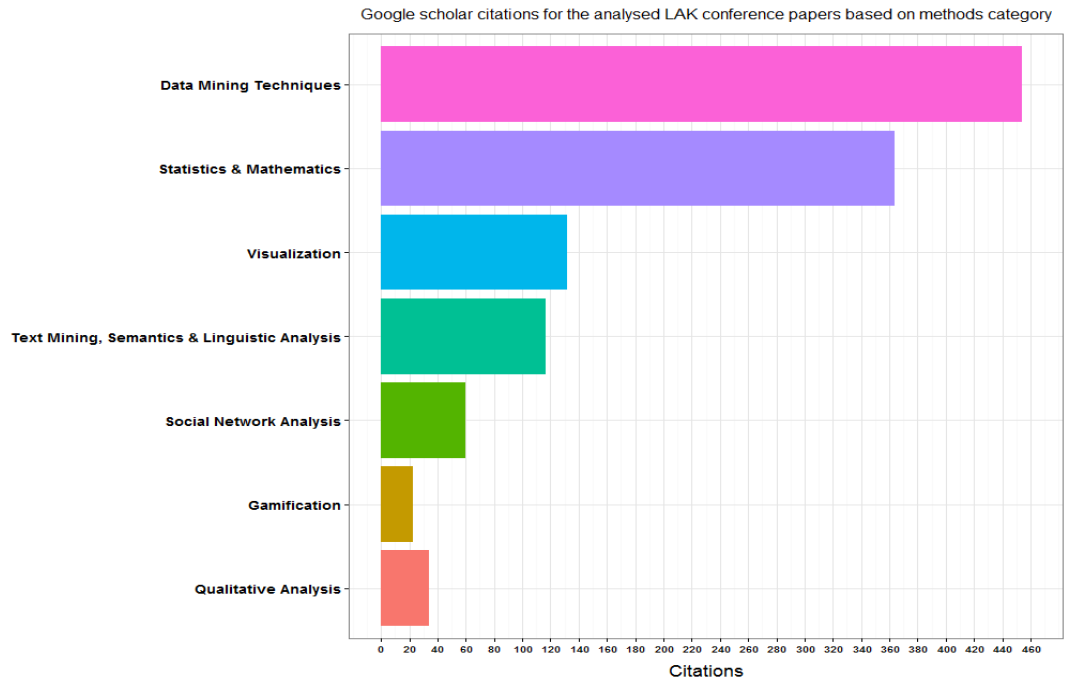

**Figure 3. Density plot of citations grouped by year**

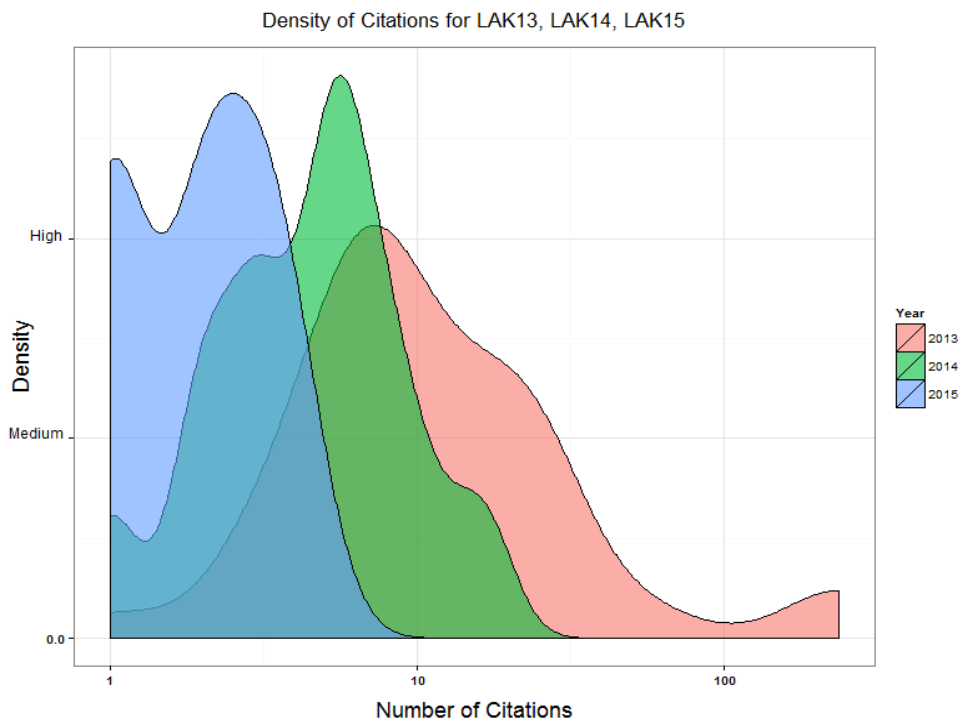

Some of 2013 publications attracted numerous citations which exceed the expectations such as (Kizilcec, Piech & Schneider, 2013; Pardos et al., 2013). A descriptive analysis of the articles in 2013 leads to: median=8, mean=24.37, max=236; articles in 2014: median=5, mean=5.56, max=17; and articles of 2015: median =1, mean=1.42, max=4. The low number of citations for 2015 is reasonable as the time span between this survey and 2015 LAK publication is around six months.

**Challenges & Constraints**

The data collection and analysis through learning analytics methods lead to questions related to ownership, privacy and ethical issues (Khalil & Ebner 2015; Khalil & Ebner, 2016). We summarize our experience and previous research studies as the following: A) Privacy, in which the learning analytics specialists need to carefully deliberate the potential privacy issues while collecting, analyzing, and intervene of the students' data. B) Transparency of disclosing information of learners and the needs to proclaim consent by the students. C) Assuring security and achieving the CIA which is an acronym that refers to Confidentiality, Integrity and Accessibility such as storing of learners records. D) The ownership of the collected and analysed data.

**Conclusion**

Learning Analytics is a promising area which provides the adequate tools and methods to optimize the learning mechanism in the different environments of educational technology platforms. In this paper, we did a meta-analysis study on publications in the last two years and classified seven different categories of techniques that have been used in that period. We noticed that learning analytics researchers adopt data mining and statistics more often than other techniques. Additionally, 2013 was a stimulating year in showing MOOCs as a desirable article by a distinct number of citations. Moreover, we also see that some publications have had a high impact on education with their peak Google scholar score. In fact, the upcoming learning analytics events might show extinction of methods and an uprising appearance or emergence of new techniques, which can be allocated in our defined categories. Finally, we summarized our experience in this field and listed some of the constraints and dilemmas that negatively affect learning analytics approaches.